# Role of Ranking Algorithms for Information Retrieval

Laxmi Choudhary[1] and Bhawani Shankar Burdak[2]

[1]Banasthali University, Jaipur, Rajasthan
laxmi.choudhary23@gmail.com
[2]BIET, Sikar, Rajasthan
bhawanichoudhary92@gmail.com

## Abstract

*As the use of web is increasing more day by day, the web users get easily lost in the web's rich hyper structure. The main aim of the owner of the website is to give the relevant information according their needs to the users. We explained the Web mining is used to categorize users and pages by analyzing user's behavior, the content of pages and then describe Web Structure mining. This paper includes different Page Ranking algorithms and compares those algorithms used for Information Retrieval. Different Page Rank based algorithms like Page Rank (PR), WPR (Weighted Page Rank), HITS (Hyperlink Induced Topic Selection), Distance Rank and EigenRumor algorithms are discussed and compared. Simulation Interface has been designed for PageRank algorithm and Weighted PageRank algorithm but PageRank is the only ranking algorithm on which Google search engine works.*

## 1. Introduction

The World Wide Web (WWW) is rapidly growing on all aspects and is a massive, explosive, diverse, dynamic and mostly unstructured data repository. As on today WWW is the huge information repository for knowledge reference. There are a lot of challenges in the Web: Web is large, Web pages are semi structured, and Web information tends to be diversity in meaning, degree of quality of the information extracted and the conclusion of the knowledge from the extracted information. So it is important to understand and analyze the underlying data structure of the Web for efficient Information Retrieval. Web mining techniques along with other areas like Database (DB), Natural Language Processing (NLP), Information Retrieval (IR), Machine Learning etc. can be used to solve the challenges. Search engines like Google, Yahoo, Iwon, Web Crawler, Bing etc., are used to find information from the World Wide Web (WWW) by the users. The simple architecture of a search engine is shown in Figure 1. There are 3 important components in a search engine. They are Crawler, Indexer and Ranking mechanism. The crawler is also called as a robot or spider that traverses the web and downloads the web pages. The downloaded pages are sent to an indexing module that parses the web pages and builds the index based on the keywords in those pages. An index is generally maintained using the keywords. When a user types a query using keywords on the interface of a search engine, the query processor component match the query keywords with the index and returns the URLs of the pages to the user. But before showing the pages to the user, a ranking mechanism is done by the search engines to show the most relevant pages at the top and less relevant ones at the bottom.
Structure Mining then section 3 describes different-different types of page ranking algorithms for information retrieval in Web and then section 4 explains comparisons between the page ranking algorithms on the basis of some parameters and section 5 explains the simulation results and at last section 6 concludes this paper.

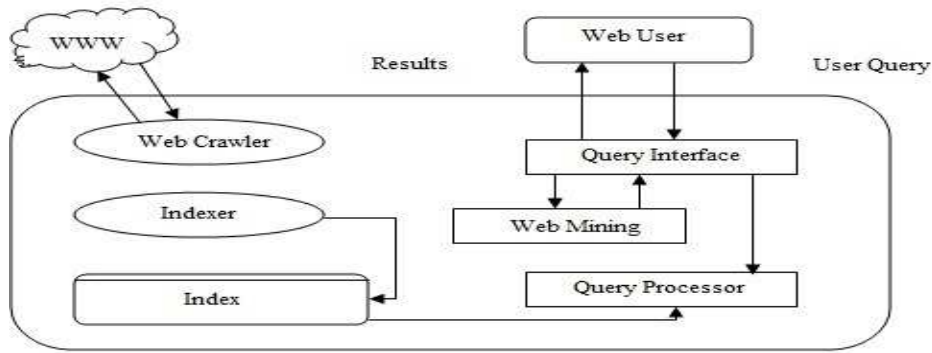

Figure 1. Simple architecture of a search engine

## 2. Related Work

Web mining is the mechanism to classify the web pages and internet users by taking into consideration the contents of the page and behavior of internet user in the past. An application of data mining technique is a web mining, which is used automatically to find and retrieve information from the World Wide Web (WWW). According to analysis targets, web mining is made of three basic branches i.e. web content mining (WCM), web structure mining (WSM) and web usage mining (WUM).

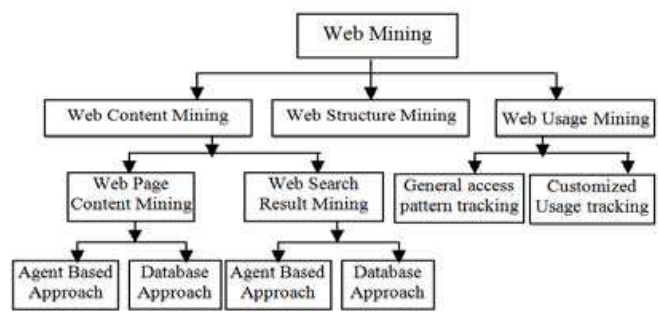

Figure 2. Web mining categories

Web Content Mining (WCM) is concerned with the retrieval of information from WWW into more structured form and indexing the information to retrieve it quickly. This is the process of extracting useful information from the contents of web documents. These web documents are collection of images, audio, video, text and structured records (such as tables and lists). This mining technique is used on the web documents and results page, which are obtained from a search engine. There are two approaches in content mining called agent based approach and database based approach. In the agent based approach, to find relevant information by using the characteristics of a particular domain while the other database approach is used to retrieve the semi-structure data from the web. Sometimes information retrieval (IR) and natural language processing (NLP) research activities uses this approach. In web content mining the relevance can be measured in this with respect to any of the following criteria such as document relevance, Query based relevance and user based role/task based relevance. Some applications of the web content mining are to identify the topics, categorize web documents, from different servers to find similar web pages, relevance applications, task based relevance applications, to provide list of relevant documents from collection and filters these documents on the basis of relevance score.

Web Usage Mining (WUM) is responsible for recording the user profile and user behavior inside the log file of the web. Web usage mining process is used to extract useful information from the data which is derived by the user while surfing on the Web. It extracts data stored in server access logs, referrer logs, agent logs, client-side cookies, user profile and meta-data. The web usage mining is an attractive technology for government agencies to find threats and fight against terrorism and using this technology, e-commerce marketing value also increases high in market and for society purposes to identify criminal activities. Also by using this technology, a better customer relationship can establish between a customer and company, so according to customer's need and demand company provides better and faster solutions.

Web Structure Mining (WSM) generates the structural summary about the Web site and Web page. It tries to discover the link structure of the hyperlinks in inter documents level. So web structure mining categorizes the web pages on the basis of the hyperlink and finds the similarity and relationship of information between different Web sites. This type of mining can be performed at intra-page or at inter-page (hyperlink level). It is important to understand the web data structure for information retrieval. Web structure mining is a challenging task to handle with the structure of the hyperlinks within the Web. There are many application areas for this new research. WSM explores the structure of the link inside the hyperlink between different documents and classify the pages of web. The number of out links i.e. links from a page and the number of in links i.e. links to a page are very important parameter in the area of web mining. The popularity of the web page is generally measured by the fact that a particular page should be referred by large number of other pages and the importance of web pages may be adjudged by a large number of out links contained by a page. So WSM becomes a very important area to be researched in the field of web mining. Figure 2 shows the general categories of web mining. In earlier days, the research was done in link analysis algorithm. So, with the growing interest in web mining, the research of structure analysis had increased and these efforts had resulted in a newly emerging research area called link mining, which is located at the intersection of the work in link analysis, hypertext and web mining, relational learning and inductive logic programming, and graph mining. Link mining had produced some agitation on some of the traditional data mining tasks. So web structure mining can use some tasks of link mining. Two basic algorithms that have been proposed to lead with those potential correlations: page ranking algorithms i.e. Google PageRank proposed by Brin and Page in 1998 and Kleinberg's hypertext induced topic selection (HITS) algorithm proposed by Kleinberg in 1998. These are used successfully and traditionally in the area of web structure mining. Both of these algorithms give equal weights to all links for deciding the rank score. So there can be different forms of link structure of the Web to be used. The aim of this technique is to make good use of understanding of these internal social assemblies of the web for link mining in applications development.

## 3. Page Ranking Algorithms

With the growing number of Web pages and users on the Web, the number of queries submitted to the search engines are also growing rapidly day by day. Therefore, the search engines needs to be more efficient in its processing way and output. Web mining techniques are employed by the search engines to extract relevant documents from the web database documents and provide the necessary and required information to the users. The search engines become very successful and popular if they use efficient ranking mechanisms. Now these days it is very successful because of its PageRank algorithm. Page ranking algorithms are used by the search engines to present the search results by considering the relevance, importance and content score and web mining techniques to order them according to the user interest. Some ranking algorithms depend only on the link structure of the documents i.e. their popularity scores (web structure mining), whereas others look for the actual content in the documents (web content mining), while some use a combination of both i.e. they use content of the document as well as the link structure to assign a rank value for a given document. If the search results are not displayed according to the user interest then the search engine will lose its popularity. So the ranking algorithms become very important. Some of the popular page ranking algorithms or approaches are discussed below.

### 3.1 PageRank Algorithm

PageRank algorithm is developed by Brin and Page during their Ph. D at Stanford University based on the citation analysis. PageRank algorithm is used by the famous search engine that is Google. This algorithm is the most commonly used algorithm for ranking the various pages.

Working of the PageRank algorithm depends upon link structure of the web pages. The PageRank algorithm is based on the concepts that if a page contains important links towards it then the links of this page towards the other page are also to be considered as important pages. The PageRank considers the back link in deciding the rank score. If the addition of the all the ranks of the back links is large then the page then it is provided a large rank. Therefore, PageRank provides a more advanced way to compute the importance or relevance of a web page than simply counting the number of pages that are linking to it. If a backlink comes from an important page, then that backlink is given a higher weighting than those backlinks comes from non-important pages. In a simple way, link from one page to another page may be considered as a vote. However, not only the number of votes a page receives is considered important, but the importance or the relevance of the ones that cast these votes as well. We assume page A has pages $T_1...T_n$ which point to it i.e., are links. The variable d is a damping factor, which value can be set between 0 and 1. We usually set the value of d to 0.85. $PR(T_1)$ is the incoming link to page A and $C(T_1)$ is the outgoing link from page $T_1$ ( such as $PR(T_1)$). The PageRank of a page A is given by the following (1):

$$PR(A) = (1-d) + d(PR(T_1)/C(T_1) +...+PR(T_n/C(T_n))) \qquad (1)$$

The damping factor is used to stop the other pages having too much influence; this total vote is damped down by multiplying it by 0.85. One important thing is noted that the page ranks form a probability distribution over web pages, so the sum of all web pages' page ranks will be one and the d damping factor is the probability at each page the random surfer will get bored and request another random page. Another simplified version of PageRank is given by:

$$PR(N) = \sum_{m \in Bn} PR(M)/L(M) \qquad (2)$$

Where the page rank value for a web page u is dependent on the page rank values for each web page v out of the set $B_n$ (This set contains all pages linking to web page N), divided by the number *L (M)* of links from page M. An example of back link is shown in figure 3 below. N is the back link of M & Q and M & Q are the back links of O.

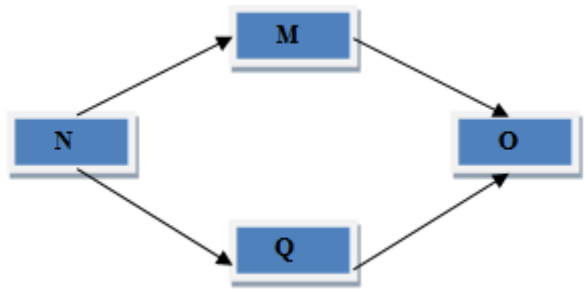

Figure 3. Back links example

Let us take an example of hyperlink structure of four pages *A, B, C* and *D* as shown in Figure 4. The PageRank for pages *A, B, C* and *D* can be calculated by using (1).

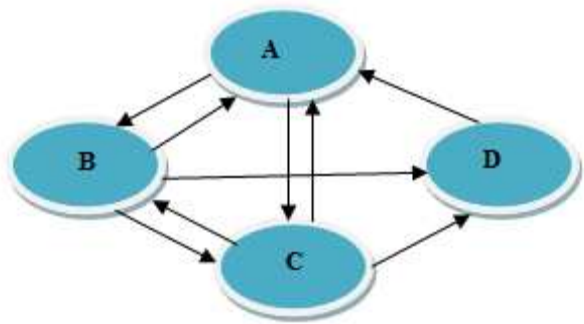

Figure 4. Hyperlink structure of four pages

Let us assume the initial PageRank as 1 and do the calculation. The value of damping factor d is put to 0.85.

$PR(A)$ = (1-d) + d (PR(B)/C(B)+PR(C)/C(C)+PR(D)/C(D))
    = (1-0.85) + 0.85(1/3+1/3+1/1)
    = 1.5666667                                                                              (3)
$PR(B)$ = (1-d) + d((PR(A)/C(A) + (PR(C)/C(C))
    = (1-0.85) + 0.85(1.5666667/2+1/3)
    = 1.0991667                                                                              (4)
$PR(C)$ = (1-d) + d((PR(A)/C(A) + (PR(B)/C(B))
    = (1-0.85) + 0.85(1.5666667/2+1.0991667/3)
    = 1.127264                                                                               (5)
$PR(D)$ = (1-d) + d((PR(B)/C(B) + (PR(C)/C(C))
    = (1-.085) + 0.85(1.0991666/3+1.127264/3)
    = 0.7808221                                                                              (6)

For the second iteration by taking the above *PageRank* values from (3), (4), (5) and (6). The second iteration PageRank values are as following:

$PR(A)$ = 0.15 + 0.85((1.0991667/3) + (1.127264/3)+(0.7808221/1)
    = 1.4445208                                                                              (7)
$PR(B)$ = 0.15 + 0.85((1.4445208/2) + (1.127264/3))
    = 1.0833128                                                                              (8)
$PR(C)$ = 0.15 + 0.85((1.4445208/2) + (1.0833128/3))
    = 1.07086                                                                                (9)
$PR(D)$ = 0.15 + 0.85((1.0833128/3)+(1.07086/3))
    = 0.760349                                                                              (10)

During the computation of 34th iteration, the average of the all web pages is 1. Some of the PageRank values are shown in Table 1. The table with the graph is shown in the simulation results section.

Table 1. Iterative Calculation for PageRank

| Iteration | A | B | C | D |
|---|---|---|---|---|
| 1 | 1 | 1 | 1 | 1 |
| 2 | 1.5666667 | 1.0991667 | 1.127264 | 0.7808221 |
| 3 | 1.4445208 | 1.0833128 | 1.07086 | 0.760349 |
| .. | .. | .. | .. | .. |
| .. | .. | .. | .. | .. |
| 17 | 1.3141432 | 0.9886763 | 0.9886358 | 0.7102384 |
| 18 | 1.313941 | 0.9885384 | 0.98851085 | 0.71016395 |
| 19 | 1.3138034 | 0.98844457 | 0.98842573 | 0.7101132 |

One thing is noted here that the rank of a page is divided evenly among it's out-links to contribute to the ranks of the pages. The original PageRank equation is a recursive which follows recursive process, starts with a given by default PageRank value i.e. 1 and computes the all iteration until all pages starts to repeat same PageRank values individually and at last find their average PageRank value that should be 1. PageRank can be calculated using a simple iterative method and corresponds to the principal an eigen vector of the normalized link matrix of the web. PageRank algorithm needs a few hours to calculate the rank of millions of pages and provides efficient output of millions pages.

For a small set of pages, it is easy to calculate and find out the PageRank values but for a Web having large set of pages or billions of pages, it is not easy to do the calculation like above. In the above Table 1, you can notice that PageRank of *A* is higher than PageRank of *B*, *C* and *D*. It is because Page *A* has 3 incoming links, Page *B*, *C* and *D* have 2 incoming links as shown in Figure 4. Page B has 2 incoming links and 3 outgoing link, page C has 2 incoming links and 3 outgoing links and page *D* has 1 incoming link and 2 outgoing links. From the Table 1, after the 34th iteration, the PageRank for the pages gets normalized.

## 3.2 Weighted Page Rank

Weighted PageRank Algorithm is proposed by Wenpu Xing and Ali Ghorbani. Weighted PageRank algorithm (WPR) is the modification of the original PageRank algorithm. WPR decides the rank score based on the popularity of the pages by taking into consideration the importance of both the inlinks and outlinks of the pages. This algorithm provides high value of rank to the more popular pages and does not equally divide the rank of a page among it's outlink pages. Every out-link page is given a rank value based on its popularity. Popularity of a page is decided by observing its number of in links and out links. As suggested, the performance of WPR is to be tested by using different websites and future work include to calculate the rank score by utilizing more than one level of reference page list and increasing the number of human user to classify the web pages.

The importance is assigned in terms of weight values to the incoming and outgoing links and are denoted as $W^{in}_{(m,n)}$ and $W^{out}_{(m,n)}$ respectively. $W^{in}_{(m,n)}$ as shown in equation (11) is the weight of *link*(*m, n*) computed depend on the number of incoming links of page *n* and the number of incoming links of all reference pages of page *m*.

$$W^{in}_{(m,n)} = I_n / \sum_{P \in Re(m)} I_p \qquad (11)$$

$$W^{out}_{(m,n)} = O_n / \sum_{P \in Re(m)} O_p \qquad (12)$$

Where *In* and *Ip* denote the number of incoming links with respect to page *n* and page *p*. $Re_{(m)}$ represents the all reference pages list of page *m*. Similarly computation performed for $W^{out}_{(m,n)}$ as shown in equation (12) is the weight of *link*(*m, n*) which is depend on the number of outgoing links of page *n* and the number of outgoing links of all the reference pages of *m*. Where as $O_n$ and $O_p$ are the number of outgoing links with respect to page *n* and *p*. The formula as proposed for the *WPR* is as shown in equation (13) which is a modification of the PageRank formula.

$$WPR(n) = (1-d) + d \sum_{m \in B(n)} WPR(m)\ W^{in}_{(m,n)}\ W^{out}_{(m,n)} \qquad (13)$$

WPR calculation calculated for the same hyperlink structure as shown in Figure 5. The *WPR* equation for Page *A*, *B*, *C* and *D* are as follows.

$$WPR(A) = (1-d) + d \sum WPR(B)\ W^{in}_{(B,A)}\ W^{out}_{(B,A)} + WPR(C)\ W^{in}_{(C,A)}\ W^{out}_{(C,A)} + WPR(D)\ W^{in}_{(D,A)}\ W^{out}_{(D,A)} \qquad (14)$$

So for getting the value of *WPR(A)*, before it we will calculate the value of incoming links and outgoing links weight as bellow:

$W^{in}_{(B,A)} = I_A/(I_A+I_C)$
　　　　$= 3/(3+2)$
　　　　$= 3/5$ (15)

$W^{out}_{(B,A)} = O_A/(O_A+O_C+O_D)$
　　　　$= 2/(2+3+1)$
　　　　$= 1/3$ (16)

$W^{in}_{(C,A)} = I_A/(I_A+I_B)$
　　　　$= 3/(3+2)$

$$= 3/5 \tag{17}$$

$$W^{out}_{(C,A)} = O_A/(O_A+O_B+O_D)$$
$$= 2/(2+3+1)$$
$$= 2/6$$
$$= 1/3 \tag{18}$$

$$W^{in}_{(D,A)} = I_A/(I_B+I_C)$$
$$= 3/(2+2)$$
$$= 3/4 \tag{19}$$

$$W^{out}_{(D,A)} = O_A/O_A$$
$$= 2/2$$
$$= 1 \tag{20}$$

Now these inlinks and outlinks weight, equation numbers (15, 16, 17, 18, 19, 20) are put in the equation (14) to calculate the weighted rank of the nodes *A, B, C,* and *D* as following:

$$WPR(B) = (1-d) + d \sum WPR(A) \, W^{in}_{(A,B)} \, W^{out}_{(A,B)} + WPR(C) \, W^{in}_{(C,B)} \, W^{out}_{(C,B)} \tag{21}$$

$$WPR(C) = (1-d) + d \sum WPR(A) \, W^{in}_{(A,C)} \, W^{out}_{(A,C)} + WPR(B) \, W^{in}_{(B,C)} \, W^{out}_{(B,C)} \tag{22}$$

$$WPR(D) = (1-d) + d \sum WPR(B) \, W^{in}_{(B,D)} \, W^{out}_{(B,D)} + WPR(C) \, W^{in}_{(C,D)} \, W^{out}_{(C,D)} \tag{23}$$

For *WPR(A)* calculation the value of d is set to 0.85(standard value) and the initial values of *WPR(B), WPR(C)* and *WPR(D)* is considered 1, so calculation for 1$^{st}$ iteration as follows:

$$WPR(A) = (1- 0.85) + 0.85(1* 3/5 *1/3 + 1* 3/5 *1/3 + 1* 3/4 *1)$$
$$= 1.127 \tag{24}$$

$$W^{in}_{(A,B)} = I_B/(I_B+I_C+I_D)$$
$$= 2/(2+2+2)$$
$$= 2/6$$
$$= 1/3 \tag{25}$$

$$W^{out}_{(A,B)} = O_B/(O_B+O_C)$$
$$= 3/(3+3)$$
$$= 3/6$$
$$= 1/2 \tag{26}$$

$$W^{in}_{(C,B)} = I_B/(I_A+I_B)$$
$$= 2/(3+2)$$
$$= 2/5 \tag{27}$$

$$W^{out}_{(C,B)} = O_B/(O_A+O_B+O_D)$$
$$= 3/(2+3+1)$$
$$= 3/6$$
$$= 1/2 \tag{28}$$

Again now for calculation of *WPR(B)* these equations (25, 26, 27, 28) are put in to equation (21). In this the initial value of *WPR(C)* is set to 1.

$$WPR(B) = (1- 0.85) + 0.85(1.127*1/3*1/2 + 1*2/5 *1/2)$$
$$= (0.15) + 0.85(1.127*0.33*0.50+1*0.40*0.50)$$
$$= 0.4989 \tag{29}$$

$$W^{in}_{(A,C)} = I_C/(I_B+I_C+I_D)$$
$$= 2/(2+2+2)$$
$$= 2/6$$
$$= 1/3 \tag{30}$$

$$W^{out}_{(A,C)} = O_C/(O_B+O_C)$$
$$= 3/(3+3)$$
$$= 3/6$$
$$= 1/2 \tag{31}$$

$$W^{in}_{(B,C)} = I_C/(I_A+I_B)$$
$$= 2/(3+2)$$
$$= 2/5 \tag{32}$$

$$W^{out}_{(B,C)} = O_C/(O_A+O_C+O_D)$$

$$= 3/(2+3+1)$$
$$= 3/6$$
$$= 1/2 \qquad (33)$$

By substituting the values of equations (24), (29), (30), (31), (32) and (33) to equation (22), you will get the *WPR* of Page *C* by taking *d* as 0.85.

$$WPR(C) = (1 - 0.85) + 0.85((1.127 *1/3 *1/2) + (0.499 * 2/5 *1/2))$$
$$= (0.15) + 0.85((1.127*0.33*0.50) + (0.499 * 0.40 * 0.50))$$
$$= 0.392 \qquad (34)$$

$$W^{in}_{(B,D)} = I_D/(I_B+I_C)$$
$$= 2/(2+2)$$
$$= 2/4 = 1/2 \qquad (35)$$

$$W^{out}_{(B,D)} = O_D/O_A$$
$$= 2/2$$
$$= 1 \qquad (36)$$

$$W^{in}_{(C,D)} = I_D/(I_A+I_B)$$
$$= 2/(2+3)$$
$$= 2/5 \qquad (37)$$

$$W^{out}_{(C,D)} = O_D/(O_A+O_B+O_D)$$
$$= 2/(2+3+1)$$
$$= 2/6$$
$$= 1/3 \qquad (38)$$

Again by substituting the values of equations (29), (34), (35), (36), (37) and (38) to equation (23), you will get the *WPR(D)* by taking *d* as 0.85.

$$WPR(D) = (1- 0.85) + 0.85((0.499 *1/2 *1) + (0.392 * 2/5*1/3))$$
$$= (0.15) + 0.85((0.499 *0.50 *1) + (0.392 * 0.40*0.33))$$
$$= 0.406 \qquad (39)$$

The values of *WPR(A)*, *WPR(B)*, *WPR(C)* and *WPR(D)* are shown in equations (24), (29), (34) and (39) respectively. The relation between these are *WPR(A)>WPR(B)>WPR(D)>WPR(C)*. This results shows that the Weighted PageRank order is different from *PageRank*.

For the same above example, the iterative computation of Weighted PageRank algorithm is computed. The some Weighted PageRank values are shown in Table 2. The table values with the chart are shown in the simulation results section.

So we can easily differentiate the WPR from the PageRank, categorized the resultant pages of a query into four categories based on their relevancy to the given query. They are:

> ➢ Very Relevant Pages (VR): The very relevant pages contain very important information related to a given query.

> ➢ Relevant Pages (R): The Relevant pages do not have important information about given query.

> ➢ Weak Relevant Pages (WR): The Weak Relevant Pages do not have the relevant information but may have the query keywords.

> ➢ Irrelevant Pages (IR): The Irrelevant Pages do not have both relevant information and query keywords.

Both the PageRank and WPR algorithms provide pages in the sorting order according to their ranks to users for a given query. So the order of relevant pages and their numbering are very important for users in the resultant list.

Table 2. Iterative calculation values for weighted pagerank

| Iteration | A | B | C | D |
|---|---|---|---|---|
| 1 | 1 | 1 | 1 | 1 |
| 2 | 1.1275 | 0.47972 | 0.3912 | 0.19935 |
| 3 | 0.425162 | 0.27674 | 0.25727 | 0.18026 |
| 4 | 0.355701 | 0.244128 | 0.24189 | 0.177541 |
| 5 | 0.34580 | 0.247110 | 0.239808 | 0.17719 |
| 6 | 0.34454 | 0.23957 | 0.23953 | 0.17714 |
| 7 | 0.34438 | 0.23950 | 0.23950 | 0.17714 |
| 8 | 0.34436 | 0.23950 | 0.23949 | 0.17714 |

### 3.3 HITS Algorithm

The HITS algorithm is proposed by Kleinberg in 1988. HITS algorithm identifies two different forms of Web pages called hubs and authorities. Authorities are pages having important contents. Hubs are pages that act as resource lists, guiding users to authorities. Thus, a good hub page for a subject points to many authoritative pages on that content, and a good authority page is pointed by many good hub pages on the same subject. Hubs and Authorities are shown in figure 5. In this a page may be a good hub and a good authority at the same time. This circular relationship leads to the definition of an iterative algorithm called HITS (Hyperlink Induced Topic Selection). HITS algorithm is ranking the web page by using inlinks and outlinks of the web pages. In this a web page is named as authority if the web page is pointed by many hyper links and a web page is named as hub if the page point to various hyperlinks. An Illustration of hub and authority are shown in figure 5. HITS is, technically, a link based algorithm. In HITS algorithm, ranking of the web page is decided by analyzing their textual contents against a given query. After collection of the web pages, the HITS algorithm concentrates on the structure of the web only, neglecting their textual contents. Original HITS algorithm has some problems which are given below.

(i) High rank value is given to some popular website that is not highly relevant to the given query.
(ii) Topic Drift occurs when the hub has multiple topics as equivalent weights are given to all the outlinks of a hub page.
(iii) In efficiency: graph construction should be performed on line.
(iv) Irrelevant links: Advertisements and Automatically generated links.
(v) Mutually effective relationship between hosts: on one site, multiple documents are pointing to document *D* at another site and retrieve their hub scores and the authority score of *D*.

Because of these above problems, the HITS algorithm is not preferred to be used in Google search engine. So PageRank algorithm is used in Google search engine because of its preference and efficiency.

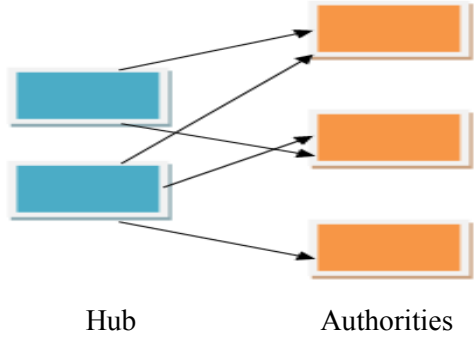

Hub            Authorities

Figure 5. Illustration of hubs and authorities

In this HITS algorithm, the hub and authority are calculated using the following algorithm.

**HITS Algorithm**

1. Initialize all weights to 1
2. Repeat until the weights converge:
3. For every hub $p \in H$
4. $Hp = \sum_{q \in Ip} Aq$
5. For every authority $p \in A$
6. $Ap = \sum_{q \in Bp} Hq$
7. Normalize

The HITS algorithm treats WWW as a directed graph $G(V,E)$, where V is a set of vertices representing pages and E is a set of edges that correspond to links. There are two main steps in the HITS algorithm. The first step is the sampling step and the second step is the iterative step. In the Sampling step, a set of relevant pages for the given query are collected i.e. a sub-graph S of G is retrieved which is high in authority pages. This algorithm starts with a root set R, a set of S is obtained, keeping in mind that S is relatively small, rich in relevant pages about the query and contains most of the good authorities. The next second step, Iterative step, finds hubs and authorities using the output of the sampling step using equations (10) and (11).

$$H_p = \sum_{q \in Ip} A_q \qquad (40)$$

$$A_p = \sum_{q \in Bp} H_q \qquad (41)$$

Where $H_p$ represents the hub weight, $A_p$ represents the Authority weight and the set of reference and referrer pages of page p denote with respect to $I(p)$ and $B(p)$. The weight of authority pages is proportional to the summation of the weights of hub pages that links to the authority page. Another one is, hub weight of the page is proportional to the summation of the weights of authority pages that hub links to. Figure 6. shows an example of the calculation of authority and hub scores.

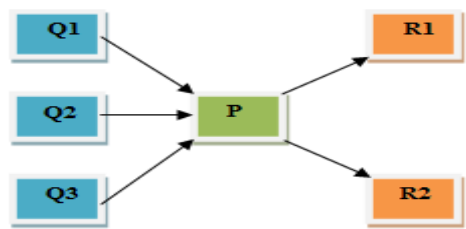

Figure 6. Calculation of hubs and authorities

From the above equations (40) and (41), the hub and authority are calculated such as:
$$A_P = H_{Q1} + H_{Q2} + H_{Q3} \tag{42}$$
$$H_P = A_{R1} + A_{R2} \tag{43}$$

### 3.4 Distance Rank Algorithm

A distance rank algorithm is proposed by Ali Mohammad Zareh Bidoki and Nasser Yazdani. This intelligent ranking algorithm based on reinforcement learning algorithm based on novel recursive method. In this algorithm, the distance between pages is considered as a distance factor to compute rank of web pages in search engine. The main goal of this ranking algorithm is computed on the basis of the shortest logarithmic distance between two pages and ranked according to them so that a page with smaller distance to assigned a higher rank. The Advantage of this algorithm is that, being less sensitive, it can find pages faster with high quality and more quickly with the use of distance based solution as compared to other algorithms. If the some algorithms provide quality output then that has some certain limitations. So the limitation for this algorithm is that the crawler should perform a large calculation to calculate the distance vector, if new page is inserted between the two pages. This Distance Rank algorithm adopts the PageRank properties i.e. the rank of each page is computed as the weighted sum of ranks of all incoming pages to that particular page. Then, a page has a high rank value if it has more incoming links on a page.

### 3.5 EigenRumor Algorithm

The EigenRumor algorithm is proposed by Ko Fujimura that ranks each blog entry on basis of weighting the hub and authority scores of the bloggers based on eigenvector calculations. So this algorithm enables a higher score to be assigned to a blog entry entered by a good blogger but not linked to by any other blogs based on acceptance of the blogger's prior work. In the recent scenario day by day number of blogging sites is increasing, there is a challenge for internet service provider to provide good blogs to the users. Page rank and HITS are very promising in providing the rank value to the blogs but some issues arise, if these two algorithms are applied directly to the blogs. These issues are:

   1. The number of links to a blog entry is generally very small. As the result, the scores of blog entries are   calculated by PageRank, for example, are generally too small to permit blog entries to be ranked by importance.

   2. Generally, some time is needed to develop a number of in- links and thus have a higher PageRank score. Since blogs are considered to be a communication tool for discussing new topics. It is desirable to assign a higher score to an entry submitted by a blogger who has been received a lot of attention in the past, even if the entry itself has no in links at first.

The rank scores of blog entries as decided by the page rank algorithm is often very low so it cannot allow blog entries to be provided by rank score according to their importance. So to resolve these issues, an EigenRumor algorithm is proposed for ranking the blogs. The EigenRumor algorithm has similarities to PageRank and HITS in that all are based on eigenvector calculation of the adjacency matrix of the links. However, in the EigenRumor model, the adjacency matrix is constructed from agent-to-object links, not page (object)-to-page (object) links. One important thing is noted that an agent is used to represent an aspect of human being such as a blogger, and an object is used to represent any object such as a blog entity. Using the EigenRumor algorithm, the hub and authority scores are calculated as attributes of agents (bloggers) and the inducement of a blog entity that does not yet have any in-link entered by the blogger can be computed.

## 4. Comparison of Various Page Rankings Algorithms

Based on the analysis, a comparison of some of various web page ranking algorithms is shown in table 3. Comparison is done on the basis of some parameters such as main technique use, methodology, input parameter, relevancy, quality of results, importance, search engine using algorithms and limitations. Among all the algorithms, PageRank and HITS are most important algorithms. PageRank is the only algorithm which is implemented in the Google search engine and HITS is used in the IBM prototype search engine called Clever. A similar algorithm is used in the other Teoma search engine and later it is used by Ask.com. The HITS algorithm can't be implemented directly in a search engine due to some problems i.e. topic drift and efficiency. That is the reason we have taken PageRank algorithm and implemented in a Java program.

Table 3. Comparison between ranking algorithms

| Algorithms / Criteria | PageRank | Weighted PageRank | HITS | Distance Rank | EigenRumor |
|---|---|---|---|---|---|
| Mining Techniques | WSM | WSM | WSM & WCM | WSM | WCM |
| Working Process | Computes values at index time and results are sorted on the priority of pages. | Computes values at index time and results are sorted on the basis of Page importance. | 'n' highly relevant pages are computed and find values on the fly. | Calculating the Minimum Average Distance Between two pages and more pages. | Use the adjacency matrix which is constructed from agent to object link not page to page. |
| I/P Parameters | Inbounds links | Inbound links and Outbound links | Inbound links and Outbound links and content | Inbounds links | Agent/Object |
| Complexity | O(log N) | <O(log N) | <O(log N) | O(log N) | <O(log N) |
| Limitations | Query independent | Query independent | Topic drift & efficiency problem | Needs to work along with PR | Used for blog ranking |
| Search Engine | Used in Google | Used in Research model | Used in IBM search engine Clever | Used in Research model | Used in Research model |

## 5. Simulation Results

The program is developed for the PageRank and Weighted PageRank algorithm using advance java language and apache tomcat server tested on an Intel Core (2 duo) with 4GB RAM machine. The input is shown in Figure 7, the user can enter the any type and any size of directed graph which contains the number of nodes, the number of incoming and outgoing links of the nodes. After press on ok1 button, matrix of entered directed graph appears beside graph on window. Now user wants the rank scores of web pages then click on submit button to calculate PageRank and Weighted PageRank comes as an output with iterative method. The output of

PageRank is shown in Figure 8(a) and PageRank values are also shown in table 4. In this simply PageRank and Weighted PageRank is calculated then their values retrieved and designed the chart of that values for web pages and compared those ranks to get higher rank web page.

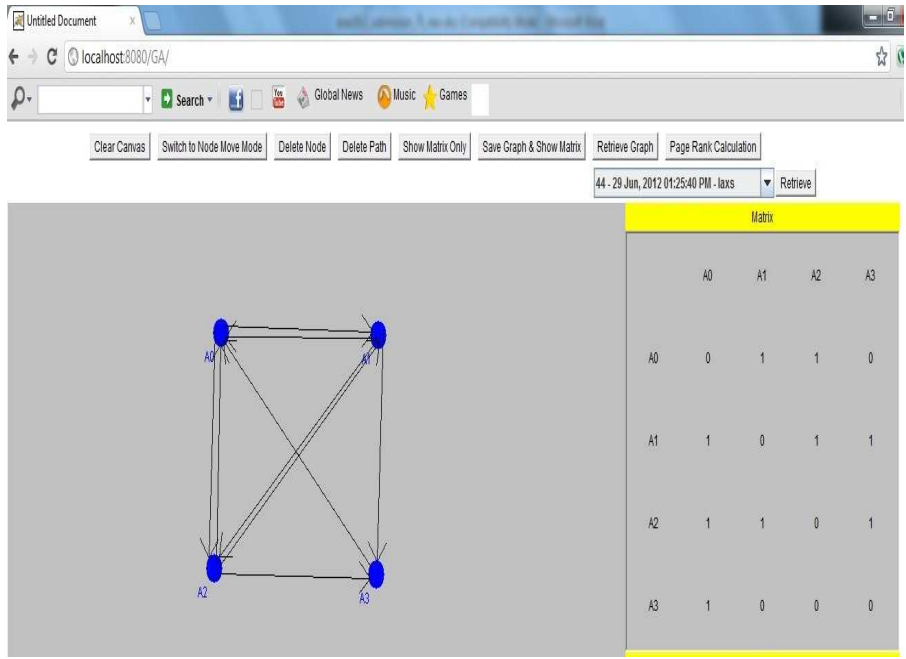

Figure 7. Simulation interface for PR program

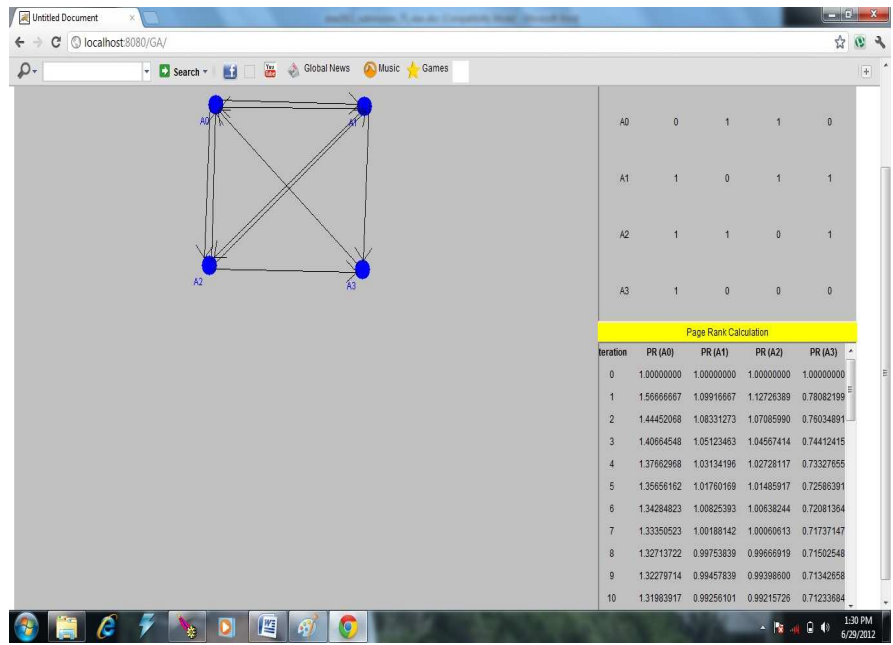

Figure 8(a). PageRank values of web pages

Table 4. Iterative Calculated Values for PageRank

| Iteration | A0 | A1 | A2 | A3 |
|---|---|---|---|---|
| 1 | 1 | 1 | 1 | 1 |
| 3 | 1.4445208 | 1.0833128 | 1.07086 | 0.760349 |
| 5 | 1.3766 | 1.0313 | 1.0272 | 0.7332 |
| 7 | 1.34284 | 1.00825 | 1.00638 | 0.720813 |
| 9 | 1.3271 | 0.9975 | 0.9966 | 0.71502 |
| 11 | 1.319839 | 0.99256 | 0.99215 | 0.712336 |
| 13 | 1.316449 | 0.990249 | 0.990061 | 0.711088 |
| 15 | 1.314874 | 0.989175 | 0.98908 | 0.710507 |
| 17 | 1.31414 | 0.988676 | 0.988635 | 0.710238 |
| 19 | 1.31380 | 0.988444 | 0.988425 | 0.710113 |

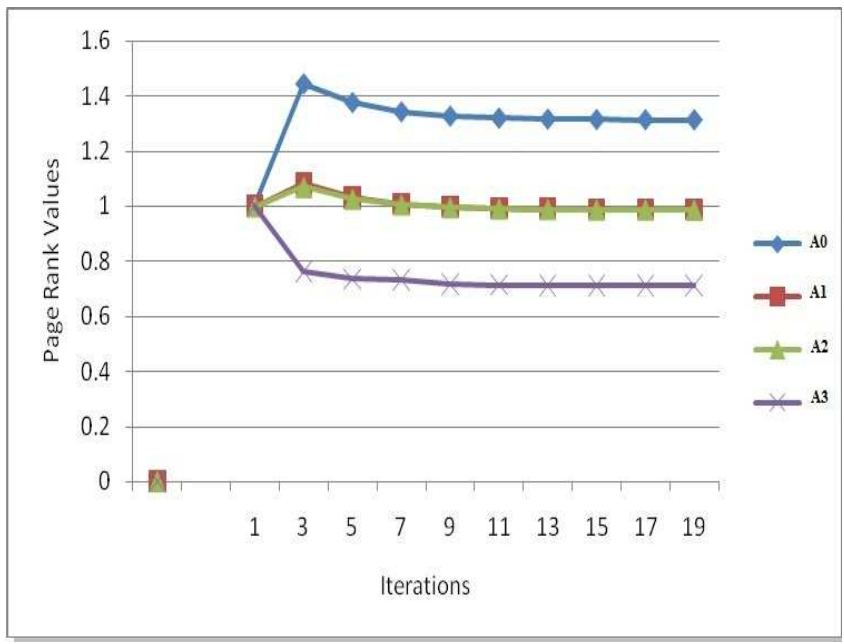

Figure 8(b). PageRank values chart with iterations

Table 5. Iterative calculated values for weighted pagerank

| Iteration | A0 | A1 | A2 | A3 |
|---|---|---|---|---|
| 1 | 1 | 1 | 1 | 1 |
| 2 | 1.1275 | 0.47972 | 0.3912 | 0.19935 |
| 3 | 0.425162 | 0.27674 | 0.25727 | 0.18026 |
| 4 | 0.355701 | 0.244128 | 0.24189 | 0.177541 |
| 5 | 0.34580 | 0.247110 | 0.239808 | 0.17719 |
| 6 | 0.34454 | 0.23957 | 0.23953 | 0.17714 |
| 7 | 0.34438 | 0.23950 | 0.23950 | 0.17714 |
| 8 | 0.34436 | 0.23950 | 0.23949 | 0.17714 |

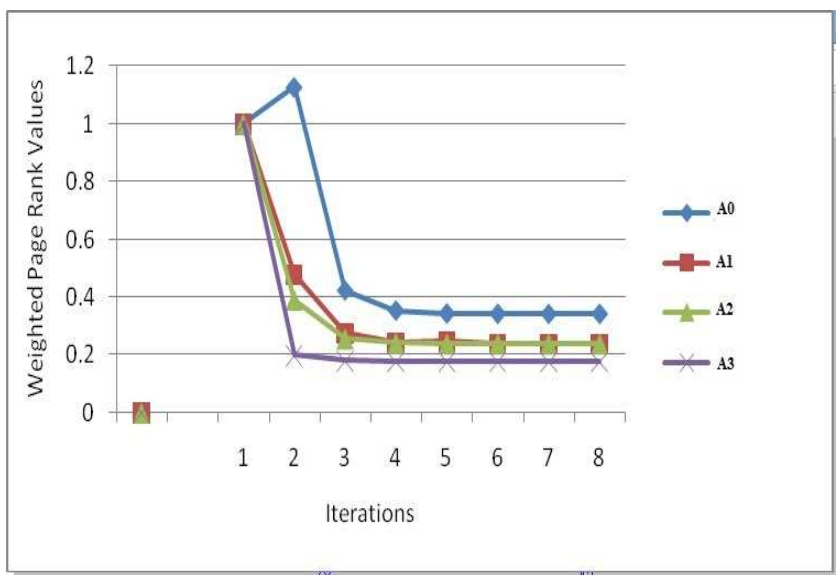

Figure 9. Weighted pagerank convergence chart

# 6. Conclusion

A typical search engine should use web page ranking techniques based on the specific needs of the users because the ranking algorithms provide a definite rank to resultant web pages. After going through this exhaustive analysis of algorithms for ranking of web pages against the various parameters such as methodology, input parameters, relevancy of results and importance of the results, it is concluded that existing algorithms have limitations in terms of time response, accuracy of results, importance of the results and relevancy of results. This paper also concludes the introduction of Web mining and the three areas of Web mining used for Information Retrieval. The main purpose is to inspect the important page ranking based algorithms used for information retrieval and compare those algorithms. An efficient web page ranking algorithm should meet out these challenges efficiently with compatibility with global standards of web technology. The work applies the PageRank program in the Web, calculates PageRank values by RageRank algorithm and weighted pagerank values using Weighted PageRank algorithm. Finally, simulation results are shown for the PageRank and Weighted PageRank algorithm and compares to web page's value in chart that shows which has higher rank values.

# 7. References

[1] N. Duhan, A. K. Sharma and K. K. Bhatia, "PageRanking Algorithms: A Survey, Proceedings of the IEEE International Conference on Advance Computing, 2009.

[2] R. Kosala, H. Blockeel, "Web Mining Research:Survey", SIGKDD Explorations, Newsletter of the ACM Special Interest Group on Knowledge Discovery and Data Mining Vol. 2, No. 1 pp 15, 2000.

[3] R. Cooley, B. Mobasher and J. Srivastava, "Web Minig: Information and Pattern Discovery on the World Wide Web," Proceedings of the 9th IEEE International Conference on Tools with Artificial Intelligence,1997.

[4] M. G. da Gomes Jr. and Z.Gong, "Web Structure Mining: An Introduction", Proceedings of the IEEE International Conference on Information Acquisition, 2005.

[5] A. Broder, R. Kumar, F Maghoul, P. Raghavan, S.Rajagopalan, R. Stata, A. Tomkins, J. Wiener, "Graph Structure in the Web", Computer Networks: The International Journal of Computer and telecommunications Networking, Vol. 33, Issue 1-6 2000.

[6] J. Kleinberg, R. Kumar, P. Raghavan, P. Rajagopalan and A. Tompkins,"Web as a Graph: Measurements, Models and methods", Proceedings of the International Conference on Combinatorics and Computing,18, 1999.

[7] L. Page, S. Brin, R. Motwani, and T. Winograd, "The Pagerank Citation Ranking: Bringing order to the Web". Technical Report, Stanford Digital Libraries SIDL-WP 1999-0120,1999.

[8] W. Xing and Ali Ghorbani, "Weighted PageRank Algorithm", Proc. Of the Second Annual Conference on Communication Networks and Services Research, IEEE.

[9] C. Ridings and M. Shishigin, "PageRank Convered". Technical Report, 2002.

[10] A. M. Zareh Bidoki and N. Yazdani, "DistanceRank:An intelligent ranking algorithm for web pages" information Processing and Management, Vol 44, No. 2, pp. 877-892, 2008.

[11] Rekha Jain, Dr G.N.Purohit, "Page Ranking Algorithms for Web Mining", International Journal of Computer application,Vol 13, Jan 2011.

[12] S. Chakrabarti, B. E. Dom, S. R. Kumar, P. Raghavan, S. Rajagopalan, A. Tomkins, D. Gibson, and J. Kleinberg, "Mining the Web's Link Structure", Computer, 32(8), PP.60–67, 1999.

[13] Lihui Chen and Wai Lian Chue, "Using Web structure and summarisation techniques for Web content mining", Information Processing and Management, Vol. 41 , pp. 1225–1242, 2005.

[14] Kavita D. Satokar and Prof.S.Z.Gawali, "Web Search Result Personalization using Web Mining", International Journal of Computer Applications, Vol. 2, No.5, pp. 29-32, June 2010.

[15] Dilip Kumar Sharma, A.k. Sharma, "A Comparative Analysis of Web Page Ranking Algorithms", International Journal on Computer Science and Engineering Vol. 02, No. 08, 2010, 2670-2676.

[16] Ko Fujimura, Takafumi Inoue and Masayuki Sugisaki,, "The EigenRumor Algorithm for Ranking Blogs", In WWW 2005 2nd Annual Workshop on the Weblogging Ecosystem, 2005.

[17] Sung Jin Kim and Sang Ho Lee, "An Improved Computation of the PageRank Algorithm", In proceedings of the European Conference on Information Retrieval (ECIR), 2002.

[18] Sergey Brin and Larry Page, "The anatomy of a Large-scale Hypertextual Web Search Engine", In Proceedings of the Seventh International World Wide Web Conference, 1998.

[19] C.. H. Q. Ding, X. He, P. Husbands, H. Zha and H. D. Simon, "PageRank: HITS and a Unified Framework for Link Analysis". 25th Annual International ACM SIGIR Conference on Research and Development in Information Retrieval, 2002.

[20] P.Boldi, M.Santini, S.Vigna, "PageRank as a Function of the Damping Factor", Proceedings of the 14th World Wide Web Conference, 2005.

[21] C.P.Lee, G.H.Golub, S.A.Zenios, A fast two-stage algorithm for computing PageRank, Technical report of Stanford University, 2003.

[22] Prasad Chebolu, Páll Melsted," PageRank and the random surfer model"; Symposium on

Discrete Algorithms Proceedings of the nineteenth annual ACM-SIAM symposium on Discrete algorithms; Pages: 1010-1018. Year : 2008.

## Authors

Laxmi Choudhary received her B.Tech degree in Computer Science and Engineering from Maharishi Arvind Institute of Engineering and Technology, Mansarowar, affiliated to Rajasthan Technical University, Kota in 2010 and M.Tech degree in Computer Science and Engineering, Banasthali Vidyapith, Jaipur Campus, affiliated to Banasthali University, Newai in 2012 respectively. Her research interests include Web Mining and it's related Algorithms, Information Retrieval and Data Mining.

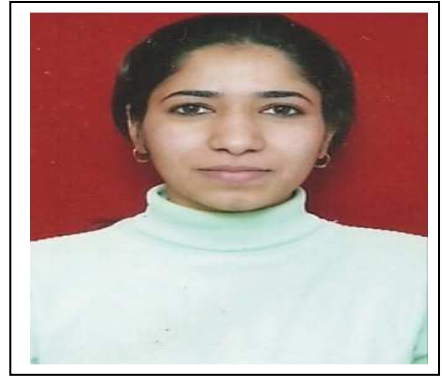

Bhawani Shankar Burdak pursuing his B.Tech degree in Information Technology from Bhartiya Institute of Engineering and technology, Sikar, affiliated to Rajasthan Technical University, Kota. His areas of interest are Database Management and Data Mining, Java Technology and Networking.

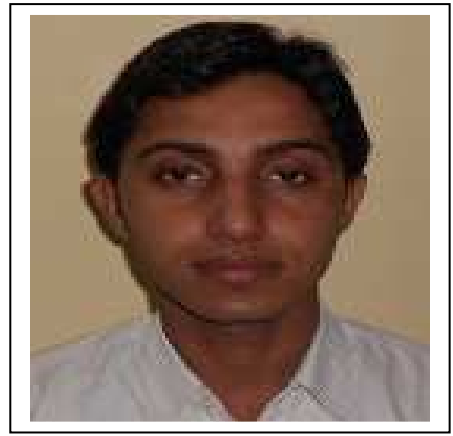